\documentclass[fleqn,12pt,twoside]{article}
\usepackage[headings]{espcrc1}

\readRCS
$Id: espcrc1.tex,v 1.2 2004/02/24 11:22:11 spepping Exp $
\usepackage{graphicx}
\usepackage[figuresright]{rotating}

\newcommand{\AmS}{{\protect\the\textfont2
  A\kern-.1667em\lower.5ex\hbox{M}\kern-.125emS}}

\title{Strangeness production from $pp$ collisions}
\author{Bing-Song Zou\address{Institute of High Energy Physics, Chinese Academy of Sciences, Beijing 100049,
China}$^,$\address[TPCSF]{Theoretical Physics Center for Sciences
Facilities, CAS, Beijing 100049, China}, Ju-Jun
Xie\address{Department of Physics, Zhengzhou University, Zhengzhou,
Henan 450001, China}$^,$\addressmark[TPCSF]}

\runtitle{Strangeness production from pp collisions}
\runauthor{Bing-Song Zou and Ju-Jun Xie}

\begin{document}

\maketitle

\begin{abstract}
The study of the strangeness production from $pp$ collisions plays
important roles in two aspects:  exploring the properties of baryon
resonances involved and understanding the strangeness production
from heavy ion collisions to explore the properties of high energy
and high density nuclear matter. Here we review our recent studies
on several most important channels for the strangeness production
from $pp$ collisions. The previously ignored contributions from
$\Delta^*(1620)$ and $N^*(1535)$ resonances are found to play
dominant role for the $pp \to n K^+ \Sigma^+$, $pp \to pK^+\Lambda$
and $pp \to pp\phi$ reactions near-thresholds. These contributions
should be included for further studies on the strangeness production
from both $pp$ collisions and heavy ion collisions.

\end{abstract}

\section{Introduction}
The strangeness production in heavy ion collisions has been proposed
to play important roles in many aspects~\cite{schne}. For an
in-depth study of the heavy ion collisions, we should, firstly, have
the proper understanding of the basic ingredients, i.e., $pp$
collisions. The study of the meson production from $pp$ collisions
itself also plays important role for exploring the baryon
spectroscopy~\cite{zou09}.

\begin{figure}[htb]
\begin{minipage}[t]{80mm}
 {\includegraphics*[scale=0.4]{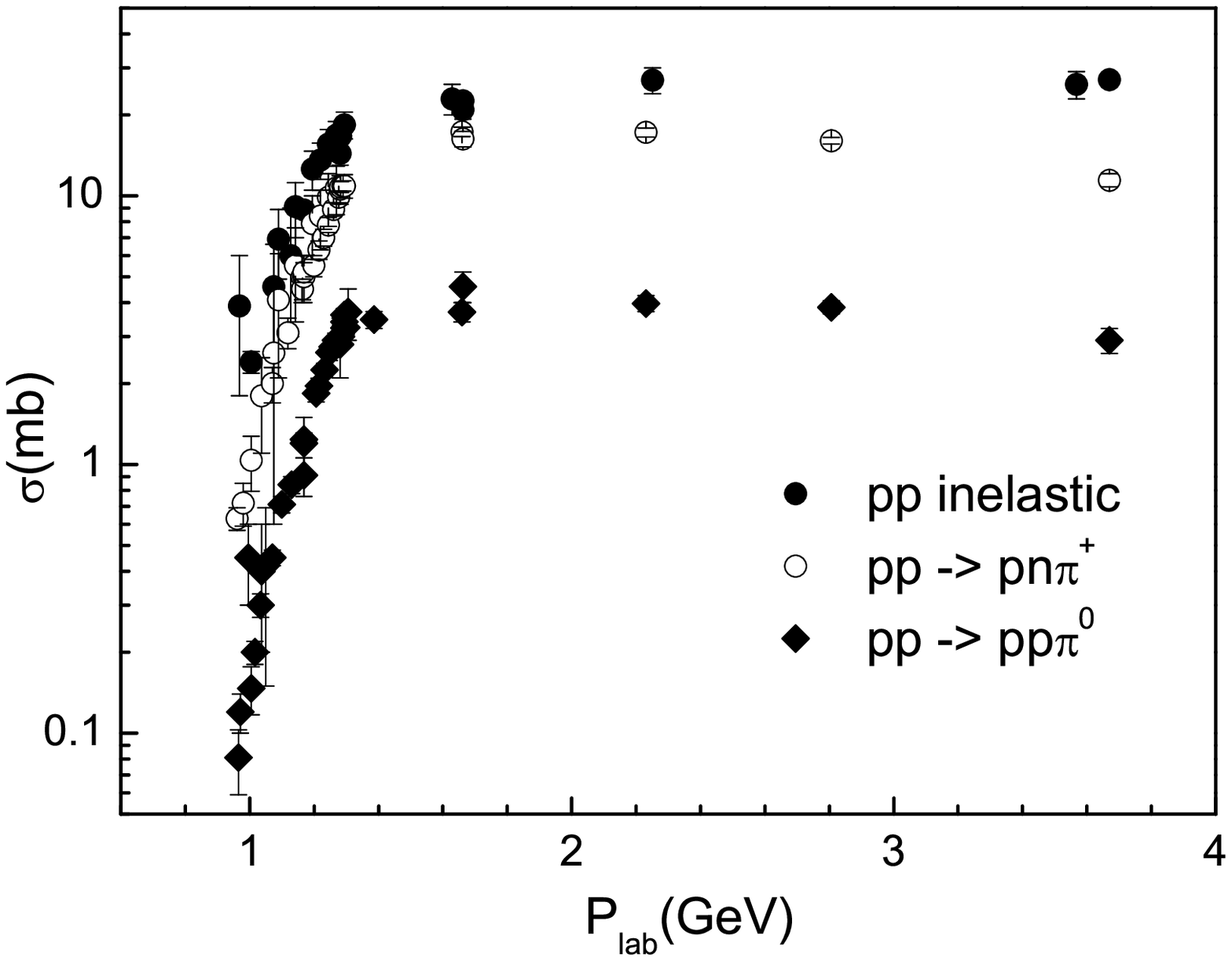}}
 \vskip -1cm
    \caption{Experimental total cross sections for the sum of the $pp$ inelastic channels,
    $pp \to p n \pi^+$, and $pp \to pp\pi^0$ reactions~\cite{pdg2008,hepdata}.}
\label{fig:pp1}
\end{minipage}
\hspace{\fill}
\begin{minipage}[t]{75mm}
{\includegraphics*[scale=0.4]{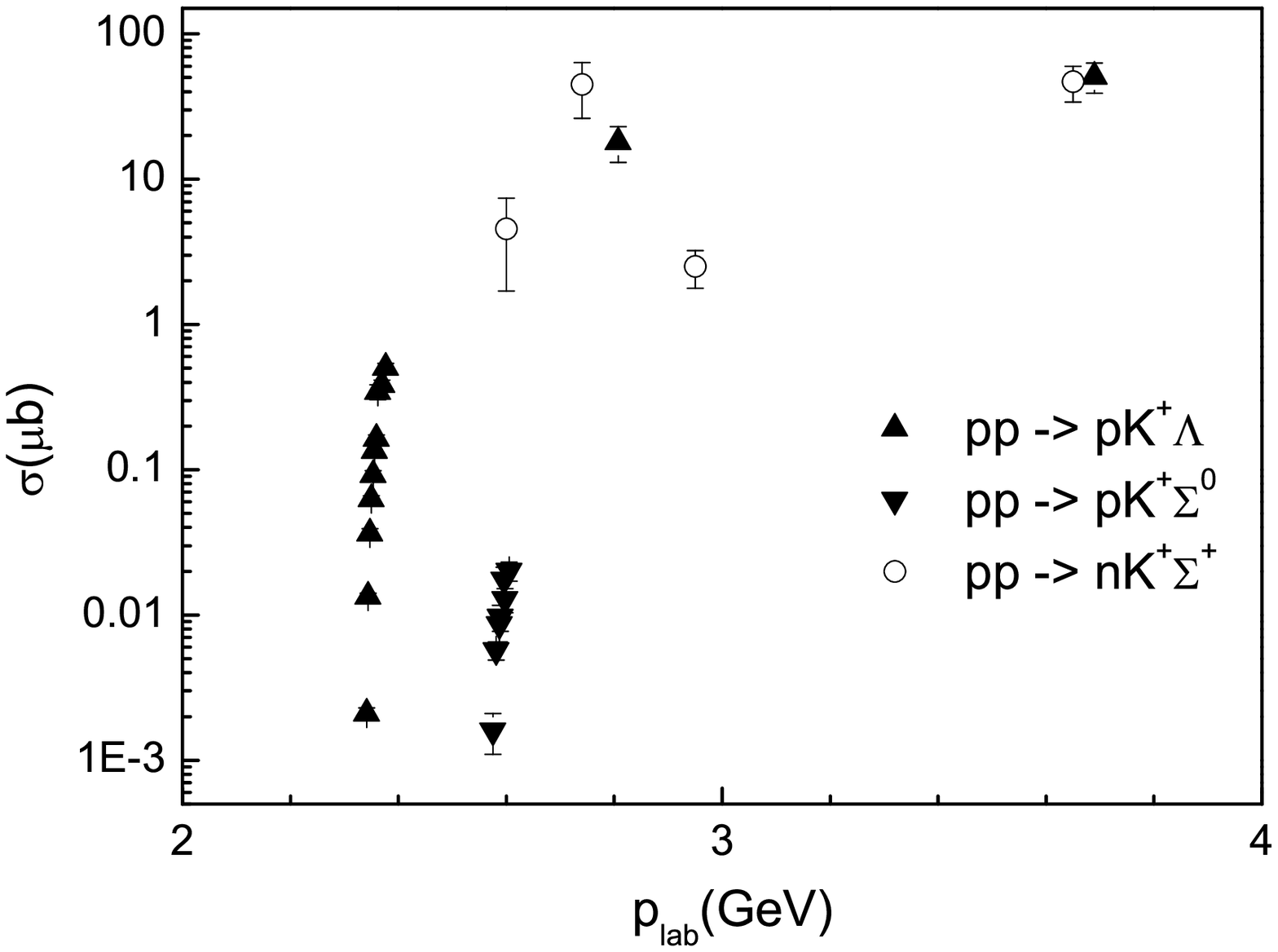}} \vskip -1cm
   \caption{Experimental total cross sections of $pp \to p
K^+\Lambda$, $pp \to p K^+ \Sigma^0$, and $pp \to nK^+\Sigma^+$
reactions~\cite{hepdata,rozek,cosyn}.} \label{fig:pp2}
\end{minipage}
\end{figure}

In the intermediate energy region, the one pion production dominates
the $pp$ inelastic scattering as shown in Fig.~\ref{fig:pp1}; while
the $pp \to p K^+\Lambda$, $pp \to p K^+ \Sigma^0$, and $pp \to
nK^+\Sigma^+$ reactions dominate the strangeness production as shown
in Fig.~\ref{fig:pp2}, for total cross
sections~\cite{pdg2008,hepdata,rozek,cosyn}. The experimental
differential cross sections for these reactions are still scarce.
Even for the largest channel, $pp \to p n\pi^+$, whether the
$N^*(1440)$ plays important role or not is still not
settled~\cite{ouyang}.

Recently, the data for the strangeness production in $pp$ collisions
at near-threshold energies have been
appearing~\cite{bale,kowina,balestra,hartman} and revealing large
discrepancy with previous theoretical predictions. Hence these
reactions have been restudied
theoretically~\cite{xiesigma,liubc,xiephi,caoxu}.

\section{Strangeness production form $pp$ collisions}

\begin{figure}[htb]
\begin{minipage}[t]{70mm}
 {\includegraphics*[scale=0.4]{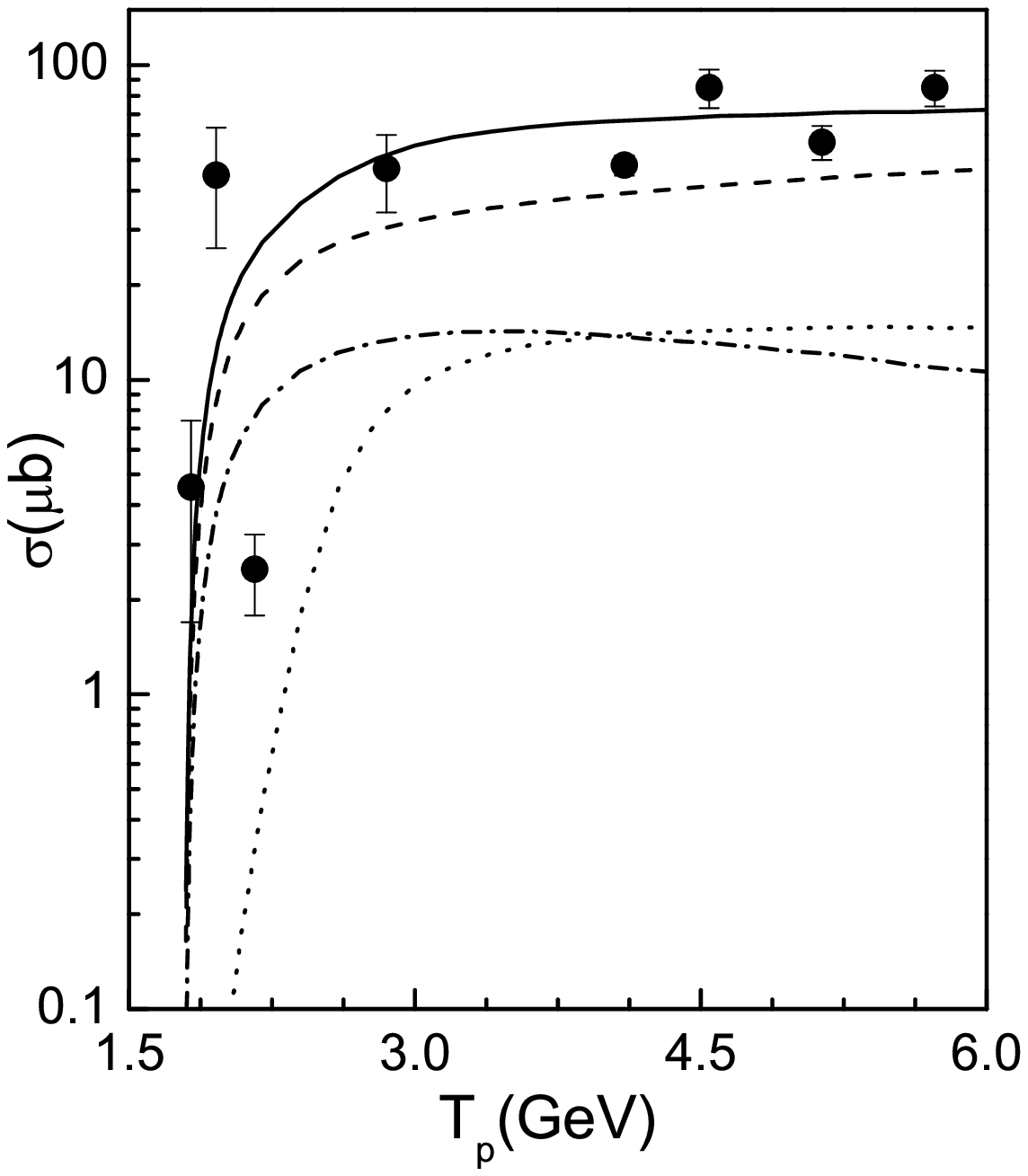}}\vskip -1cm
    \caption{Total cross section vs the beam energy $T_P$ for the $p p \to n K^+\Sigma^+$.} \label{fig:sigmaktcs}
\end{minipage}
\hspace{\fill}
\begin{minipage}[t]{70mm}
{\includegraphics*[scale=0.65]{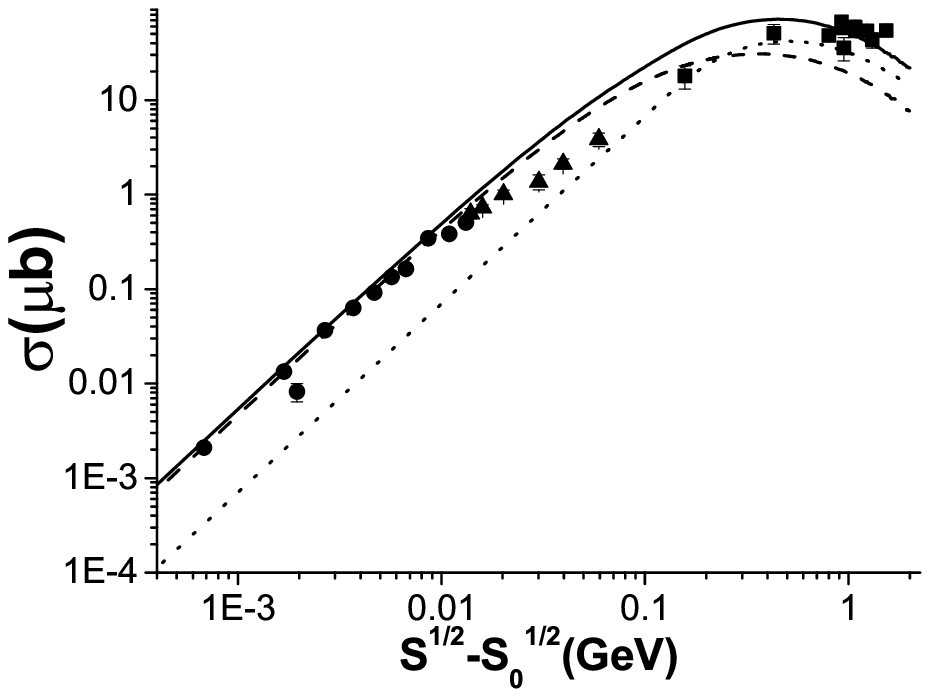}}\vskip -1cm
   \caption{Total cross section vs excess energy for the
    $pp \to p K^+ \Lambda$.} \label{fig:lktcs}
\end{minipage}
\end{figure}

The $pp \to nK^+\Sigma^+$ reaction, which has a special advantage
for absence of complication caused by $N^*$ contribution because of
the isospin and charge conversation, is a possible new excellent
source for studying $\Delta^{++*}$ resonances. The previously
theoretical works~\cite{tsushima99,gaspar} on this reaction only
reproduce the old data at higher beam energies well, but fail by
order of magnitude compared with very recent COSY-11 data at
energies close to threshold~\cite{rozek}. Recently this reaction was
restudied~\cite{xiesigma}. Besides the ingredients considered in
previous calculations, the sub-$K\Sigma$-threshold
$\Delta^{++*}(1620)$ resonance is added by taking into account both
$\pi^+$ and $\rho^+$ mesons exchange. The numerical results are
shown in Fig.~\ref{fig:sigmaktcs} together with the
data~\cite{hepdata,rozek,cosyn,baldi} for comparison. The
contributions from $\Delta^*(1620)(\pi^+$ exchange),
$\Delta^*(1620)(\rho^+$ exchange) and $\Delta^*(1920)(\pi^+$
exchange) are shown separately by dot-dashed, dashed and dotted
curves, respectively. The contribution from the $\Delta^*(1620)$
production by the $\rho^+$ exchange is found to be very important
for the whole energy range, in particular, for the two lowest data
points close to the threshold. This gives a natural source for the
serious underestimation of the near-threshold cross sections by
previous calculations \cite{tsushima99,gaspar}, which have neglected
either $\Delta^*(1620)$ resonance contribution~\cite{tsushima99} or
$\rho^+$ exchange contribution \cite{gaspar}. The solid curve in the
figure is the simple sum of the three contributions and reproduces
the COSY-11 data quite well. However, a more recent measurement of
the reaction near its threshold by ANKE collaboration~\cite{cosyn}
gives a much smaller cross section than those by
COSY-11~\cite{rozek}. Since both detectors are not $4\pi$ solid
angle detectors, there is model dependence to deduce the total cross
section from a fraction of $4\pi$ solid angle measurement. A good
Dalitz plot measurement with a good $4\pi$ solid angle detector
would be very helpful to settle down the contradiction.

It is well known that the $N^*(1535)$ resonances couples strongly to
the $\eta N$ channel~\cite{pdg2008}. Recently, it was also
found~\cite{liubc} to have strong coupling to $K\Lambda$ based on
BES data on $J/\psi \to p \bar{p} \eta$ and $J/\psi \to \bar{p} K^+
\Lambda + c.c.$ reactions~\cite{beseta}. With the large
$g_{N^*(1535)K\Lambda}$ coupling constant, the contribution from
$N^*(1535)$ to $pp \to p K^+\Lambda$ is checked ~\cite{liubc} in the
effective Lagrangian approach. The calculated results are shown in
Fig.~\ref{fig:lktcs}. The dashed and dotted curves represent the
contribution from $N^*(1535)$ and other $N^*$
resonances~\cite{tsushima}, respectively. The solid line is the sum.

\begin{figure}[htb]
\begin{minipage}[t]{70mm}
 {\includegraphics*[scale=0.3]{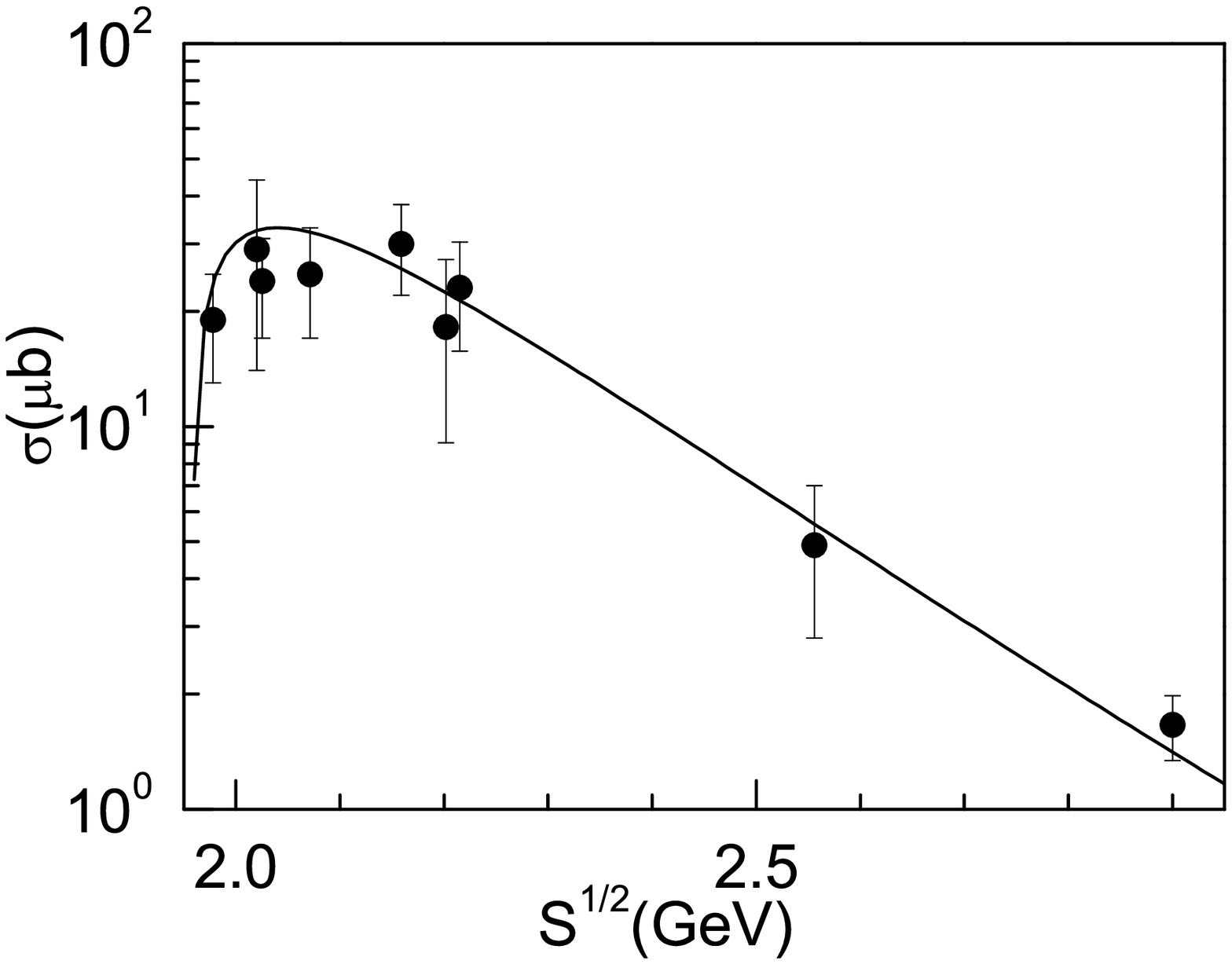}}\vskip -1cm
    \caption{Total cross section vs CM energy of the $\pi p \to n \phi$ reaction.} \label{fig:piptcs}
\end{minipage}
\hspace{\fill}
\begin{minipage}[t]{70mm}
{\includegraphics*[scale=0.3]{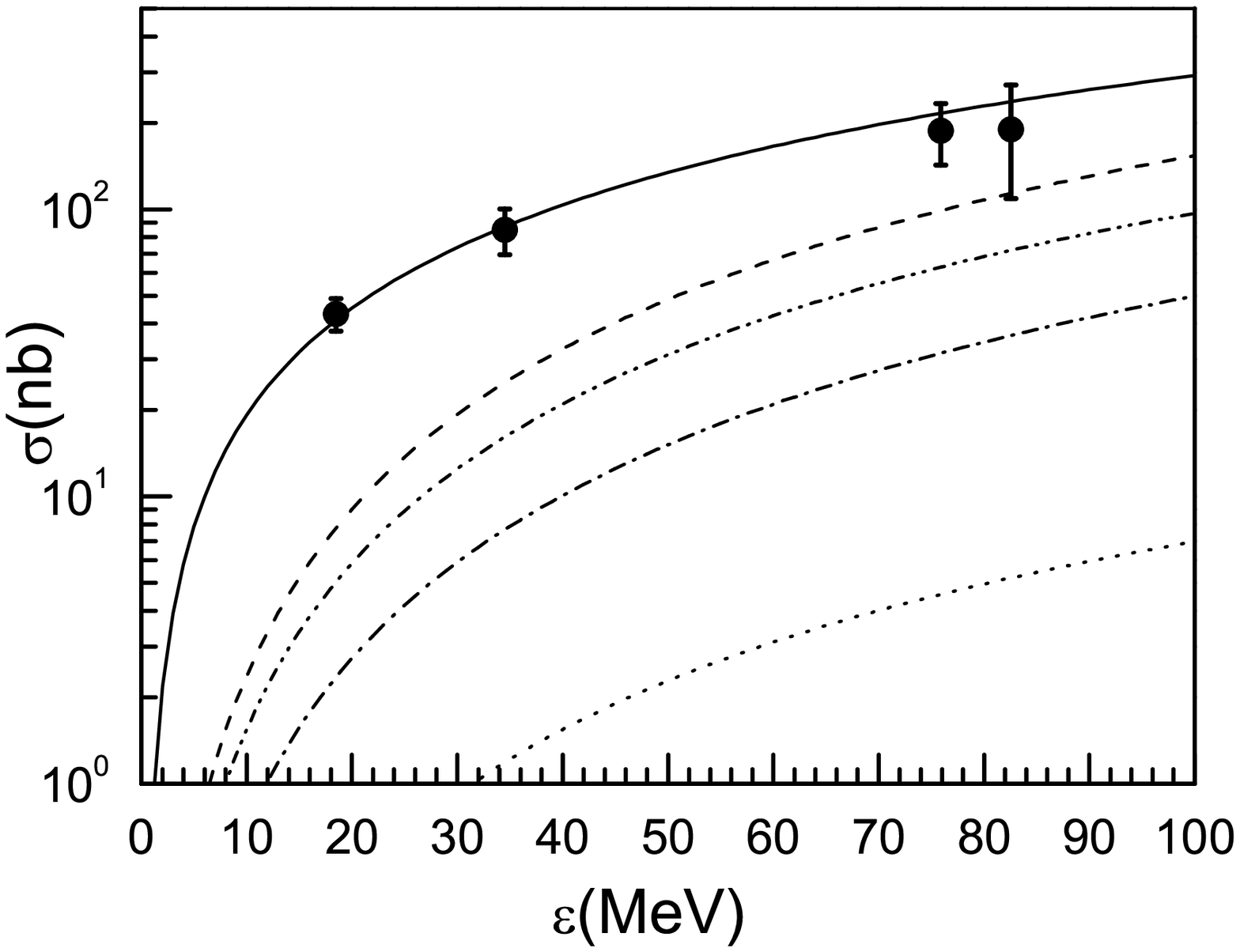}}\vskip -1cm
   \caption{Total cross section vs excess energy for the $pp \to p p \phi$ reaction.} \label{fig:pphitcs}
\end{minipage}
\end{figure}

Since the $N^*(1535)$ has strong couplings to $N\eta$, $K\Lambda$
and maybe also $N\eta^\prime$~\cite{caoxuetaprime}, there may be a
significant $s\bar{s}$ components in it. This indicates that the
$N^*(1535)$ may also have a significant coupling to the $\phi N$
channel. Assuming that the production of the $\phi$ meson in $pp$
and $\pi^- p$ collisions is predominantly through the excitation and
decay of the sub-$\phi N$-threshold $N^*(1535)$ resonance, the $pp
\to pp \phi$ and $\pi^- p \to n \phi$ reactions were
calculated~\cite{xiephi}. The results compared with the experimental
data were shown by solid curve in
Figs.~\ref{fig:piptcs}\&~\ref{fig:pphitcs}. In
Fig.~\ref{fig:pphitcs}, the double dotted-dashed, dotted,
dashed-dotted and dashed curves stand for contributions from
$\pi^0$, $\eta$, $\rho^0$-meson exchanges and their simple sum,
respectively. The solid line includes the $^1S_0$ $pp$ FSI.

\section{Summary}

In summary, the largest strangeness production channels from $pp$
collisions, $pp \to p K^+\Lambda$, $pp \to p K^+ \Sigma^0$, and $pp
\to nK^+\Sigma^+$ reactions, have been restudied theoretically by
including contributions from previously ignored $\Delta^*(1620)$ and
$N^*(1535)$ resonances~\cite{xiesigma,liubc,xiephi,caoxu}. These
sub-threshold resonances have been found to play dominant roles for
the strangeness productions in $pp$ collisions and should be
included for further studies on the strangeness production from both
$pp$ collisions and heavy ion collisions.

\section{Acknowledgments}

We would like to thank Zhen Ouyang, Bo-Chao Liu, Huan-Ching Chiang,
and Xu Cao for Collaborations on some relevant issues reviewed in
the present work. This work is partly supported by the National
Natural Science Foundation of China under grants Nos. 10875133,
10847159, 10521003 and by the Chinese Academy of Sciences under
project No. KJCX3-SYW-N2.

\end{document}